\documentclass[aps,prb,showpacs,superscriptaddress,twocolumn]{revtex4}
\usepackage{times}
\usepackage{epsfig}
\usepackage{color}
\usepackage{psfrag}

\newcommand{\ba}{\begin{eqnarray}}
\newcommand{\be}{\begin{equation}}
\newcommand{\ea}{\end{eqnarray}}
\newcommand{\ee}{\end{equation}}

\begin{document}

\title{Spin amplification, reading, and writing in transport through
anisotropic magnetic molecules}

\author{Carsten Timm}
\email{ctimm@ku.edu}
\affiliation{Department of Physics and Astronomy, University of Kansas,
Lawrence, KS 66045, USA}
\affiliation{Institut f\"ur Theoretische Physik, Freie Universit\"at Berlin,
Arnimallee 14, 14195 Berlin, Germany}
\author{Florian Elste}
\email{felste@physik.fu-berlin.de}
\affiliation{Institut f\"ur Theoretische Physik, Freie Universit\"at Berlin,
Arnimallee 14, 14195 Berlin, Germany}

\date{November 11, 2005}

\begin{abstract}
Inelastic transport through a single magnetic molecule
weakly coupled to metallic leads is studied theoretically. We consider
dynamical processes that are relevant for writing, storing, and reading spin
information in molecular memory devices. Magnetic anisotropy is found to be
crucial for slow spin relaxation. In the presence of
anisotropy we find giant spin amplification: The spin accumulated in the
leads if a bias voltage is applied to a molecule prepared in a
spin-polarized state can be made exponentially large in a characteristic energy
divided by temperature. For one ferromagnetic
and one paramagnetic lead the molecular spin can be reversed by applying a
bias voltage even in the absence of a magnetic field. We propose schemes for
reading and writing spin information based on our findings.
\end{abstract}


\maketitle

\section{Introduction}

Recently, a number of fascinating experiments have
studied transport through magnetic molecules. The current through a molecule
weakly coupled to metallic leads shows steps as a function of bias voltage
when additional molecular transitions become energetically available. The
differential conductance $dI/dV$ then shows peaks. Zeeman splitting of the
energy levels and of the peaks in $dI/dV$ has been observed for magnetic
molecules.\cite{PPG02,LSB02} Magnetic anisotropy partially lifts the
degeneracy of molecular levels, leading to fine structure of the peaks even
for vanishing magnetic field, which has been observed for $\mathrm{Mn}_{12}$
complexes.\cite{HGF05} Some of the fine structure peaks show negative
differential conductance (NDC).\cite{HGF05} The Kondo effect has also been
found for single molecules.\cite{PPG02,LSB02} The Zeeman splitting\cite{ElT05a}
as well as spin blockade and NDC\cite{RWS05,HGF05} in
magnetic molecules have been investigated theoretically. Transport through
magnetic molecules has also been studied in the STM geometry, where the
coupling to the substrate is much stronger.\cite{DuW02}

Magnetic molecules are interesting for applications that combine
spintronics,\cite{spintronics,spin2} i.e., the idea to use the electron spin in
electronic devices, with molecular electronics, i.e., the utilization of
molecules as electronic components.\cite{Joach,molel} We study processes that
are relevant for the utilization of magnetic molecules as memory cells, which
has been discussed at least since the early
1990s.\cite{Kahn,Sessoli,Joach,Werns} In this context it is crucial that the
stored information \emph{persists} over sufficiently long times and that one can
\emph{read out} and \emph{write} the information. The read out process would
naturally utilize the interaction of electrons tunneling through the molecule
with its local spin. The works mentioned above lay the ground for a study of
this aspect. On the other hand, the other two aspects---persistence and
writing---have received little attention.

In the present paper, we consider current-induced spin relaxation and
propose schemes for reading and writing the molecular spin electronically.
We show that the spin transmitted from one lead to the other before the
molecular spin relaxes can be made very large. This interesting effect of
\emph{spin amplification} forms the basis of our proposed read-out
mechanism. \emph{Magnetic anisotropy} is a key ingredient for this effect
and for molecular memory devices in general. Since in the absence of a
magnetic field the spin polarization vanishes in the steady state, we expect
any spin polarization to decay in time. For memory applications this
relaxation should be slow. That this can be accomplished by an easy-axis
anisotropy, which introduces an \emph{energy barrier} for a change of spin
direction,\cite{HGF05} has been shown for $\mathrm{Fe}_8$ complexes
by Barra \textit{et al}.\cite{BDG96}
Typical molecules we have in mind are planar
complexes of magnetic ions, e.g., porphyrin complexes like heme.

\section{Theory}

We consider a magnetic molecule with one orbital coupled
to a local spin $\mathbf{S}$, described by the Hamiltonian
\ba
H_{\mathrm{mol}} & = & (\epsilon-eV_g)\, \hat n
  + \frac{U}{2}\, \hat n(\hat n-1) - J\, \mathbf{s}\cdot\mathbf{S} \nonumber
  \\
& & {}- K_2\, (S^z)^2
  - B (s^z+S^z) .
\label{1.H2}
\ea
Here, $\hat n \equiv c^\dagger_\uparrow c_\uparrow + c^\dagger_\downarrow
c_\downarrow$ is the number operator ($c^\dagger_\sigma$ creates an electron
in the orbital) and $\mathbf{s} \equiv \sum_{\sigma\sigma'} c^\dagger_\sigma
(\mbox{\boldmath$\sigma$}_{\sigma\sigma'}/2) c_{\sigma'}$ is the
corresponding spin operator. $\epsilon$ is the
single-electron energy in the orbital, which can be shifted by the gate voltage
$V_g$, $U$ is the Coulomb repulsion between two electrons in the orbital, and
$J$ is the exchange interaction between an electron in the orbital and the local
spin of length $S\ge 1$. Easy-axis  magnetic anisotropy is assumed
($K_2>0$) for the local spin. We define the total spin
operator $\mathbf{S}_{\mathrm{tot}} \equiv \mathbf{s} + \mathbf{S}$.

The Hamiltonian in Eq.~(\ref{1.H2}) is the simplest one with an
anisotropy-induced energy barrier and already exhibits the essential physics of
reading, storing, and writing molecular-spin information, as we show in this
paper. Our model does not allow for magnetic tunneling through the energy
barrier in the absence of electron tunneling involving the leads. This effect is
included in Ref.~\onlinecite{HGF05} but is assumed to be weak there. Our model
is valid as long as the typical rate of spontaneous magnetic tunneling is small
compared to typical electronic tunneling rates.

The molecule is weakly coupled to two leads L (left) and R (right) with
Hamiltonians $H_\alpha = \sum_{\mathbf{k}\sigma}
\epsilon_{\alpha\mathbf{k}\sigma}\, a^\dagger_{\alpha\mathbf{k}\sigma}
a_{\alpha\mathbf{k}\sigma}$ where $a^\dagger_{\alpha\mathbf{k}\sigma}$ creates
an electron in lead $\alpha=\mathrm{L},\mathrm{R}$. The hybridization of the
molecular orbital with the leads is described by Hamiltonians
$H_{\mathrm{hyb},\alpha} = \sum_{\mathbf{k}\sigma} (t_\alpha\,
a^\dagger_{\alpha\mathbf{k}\sigma} c_\sigma + \mathrm{h.c.})$. In the following
we consider the case of symmetric contacts, $t_{\mathrm{L}}=t_{\mathrm{R}}$, to
avoid unnecessary complications that have nothing to do with the physics we are
interested in. However, it should be noted that asymmetry of the contacts,
which is expected to be present, can have a significant effect on transport
properties.\cite{Bul00}

The unperturbed Hamiltonian $H_{\mathrm{mol}}$ is easily diagonalized. The
eigenstates fall into sectors with $n=0,1,2$ electrons. Since
$[S_{\mathrm{tot}}^z,H_{\mathrm{mol}}]=0$ the eigenvalue $m$ of
$S_{\mathrm{tot}}^z$ is another good quantum number. For $n=0$, $2$ we
obtain $2(2S+1)$ eigenstates $\{|0,m\rangle, |2,m\rangle\}$ with energies
$\epsilon(0,m) = -K_2 m^2 - B m$ and $\epsilon(2,m) = 2(\epsilon-eV_g) + U
-K_2 m^2 - B m$, respectively.
For $n=1$ we find the eigenstates,
in terms of spin states of the electron and the local spin,
\ba
\lefteqn{ |1,m\rangle^\pm =
  \mp \frac{\sqrt{2\Delta E \mp (2K_2-J)m}}{2\sqrt{\Delta E}}\,
  |\downarrow\rangle
  |m+1/2\rangle } \nonumber \\
& & {}+ \frac{J\sqrt{S(S+1)-m^2+1/4}}{2\sqrt{\Delta E}
  \sqrt{2\Delta E\mp (2K_2-J)m}}\, |\uparrow\rangle
  |m-1/2\rangle ,
\label{C2.mixst4}
\ea
with energies
\be
\epsilon^\pm(1,m) = \epsilon - eV_g -Bm+\frac{J}{4}
  -K_2 \Big(m^2+\frac{1}{4}\Big) \pm \Delta E(m)
\label{C2.mixst5}
\ee
where $\Delta E(m) \equiv [{K_2(K_2-J)m^2+({J}/{4})^2 (2S+1)^2}]^{1/2}$.
There are two orthogonal states $|1,m\rangle^\pm$ for $-S+1/2\le m\le
S-1/2$. For the fully polarized states $|1,-S-1/2\rangle$ and
$|1,S+1/2\rangle$ the upper (lower) sign applies
if $K_2-J/2$ is positive (negative).

The hybridization is treated as a perturbation. The derivation follows
Refs.~\onlinecite{ElT05a,MAM04,KOO04,KoO05} and is not repeated here.
We obtain coupled rate equations
\be
\dot P^i = \sum_{j\neq i} P^j R_{j\to i} - P^i \sum_{j\neq i} R_{i\to j}
\label{C2.rateq}
\ee
for the probabilities $P^i$ of the molecular many-body states $|i\rangle$.
The steady state $P_0^i$ is the solution for $\dot P^i=0$.

The transition rates $R_{i\to j}$ are written as a sum over the two leads
and the two spin directions, $R_{i\to j}=\sum_{\sigma\alpha}
R^{\sigma\alpha}_{i\to j}$, with
\ba
R^{\sigma\alpha}_{i\to j} & = & \frac{1}{\tau_0} \big[
  f(\epsilon_j-\epsilon_i+e s_\alpha V/2)\, |C_{ij}^\sigma|^2
  \nonumber \\
& & {}+ f(\epsilon_j-\epsilon_i-e s_\alpha V/2)\, |C_{ji}^\sigma|^2 \big] ,
\label{C1.R4}
\ea
where $s_{\mathrm{L},\mathrm{R}}=\pm 1$, $1/\tau_0 \equiv 2\pi\,
|t|^2\, Dv_{\mathrm{uc}}/\hbar$ is the typical transition rate in terms of
the density of states $D$ (for one spin direction)
of the leads and their
unit-cell volume $v_{\mathrm{uc}}$, $V$ is the bias voltage,
$\epsilon_i$ is the energy of state $|i\rangle$, $f(x)$ is the Fermi
function, and $C_{ij}^\sigma \equiv \langle i|c_\sigma|j\rangle$.

The occupation and spin in a state
described by probabilities $P^i$ are given by $n =
\sum_i n_i P^i$ and $m = \sum_i m_i P^i$, respectively. Here, $n_i$ ($m_i$)
is the occupation (polarization) in state $|i\rangle$.
The current through lead $\alpha$ is
\be
I^\alpha =
  -e\,s_\alpha \sum_{ij} (n_i-n_j)\, \big( R^{\uparrow\alpha}_{j\to i}
  + R^{\downarrow\alpha}_{j\to i} \big)\, P^j ,
\label{C1.I5}
\ee
where we count currents from left to right as positive. The
\emph{steady-state}
currents $\langle I^\alpha\rangle$ through the two leads are equal
since the occupation is constant in time. The current scale is
$e/\tau_0$, where the rate $1/\tau_0$ comes from the transition rates.
Comparing this to typical experimental currents of
$0.1\,\mathrm{nA}$,\cite{PPG02,c60ex} we obtain $\tau_0\sim 1.6\,\mathrm{ns}$.
The \emph{spin} current
\be
I_s^\alpha = \frac{s_\alpha}{2}\, \sum_{ij} (n_i-n_j)\,
  \big( R^{\uparrow\alpha}_{j\to i} - R^{\downarrow\alpha}_{j\to i}
  \big)\, P^j
\label{C1.Is5}
\ee
is generally nonzero if the molecule is prepared in a spin-polarized state.
The molecule relaxes exponentially towards
the steady state, which for $B=0$ is nonmagnetic.
We can thus define the \emph{total transmitted spin}
\be
\Delta S^\alpha[\mathbf{P}] = \int_0^\infty dt\, I_s^\alpha(t,\mathbf{P}) ,
\label{C1.DS2}
\ee
which depends on the probabilities $\mathbf{P}=(P^1,\ldots)$ at $t=0$.

To evaluate Eq.~(\ref{C1.DS2}) we require the time dependence of $\mathbf{P}$.
We rewrite the rate equations (\ref{C2.rateq}) in matrix form,
$\dot\mathbf{P} = A\,\mathbf{P}$, where $A$ has the components
$A_{ij} = R_{j\to i}$ for $i\neq j$ and $A_{ii} = -\sum_{k\neq i} R_{i\to k}$.
The solution is $\mathbf{P}(t) =
e^{At}\,\mathbf{P}(0)$, which allows to calculate occupation,
polarization, and (spin) current as functions of time for given initial
conditions.

\begin{figure}[t]
\centerline{\includegraphics[width=1.60in]{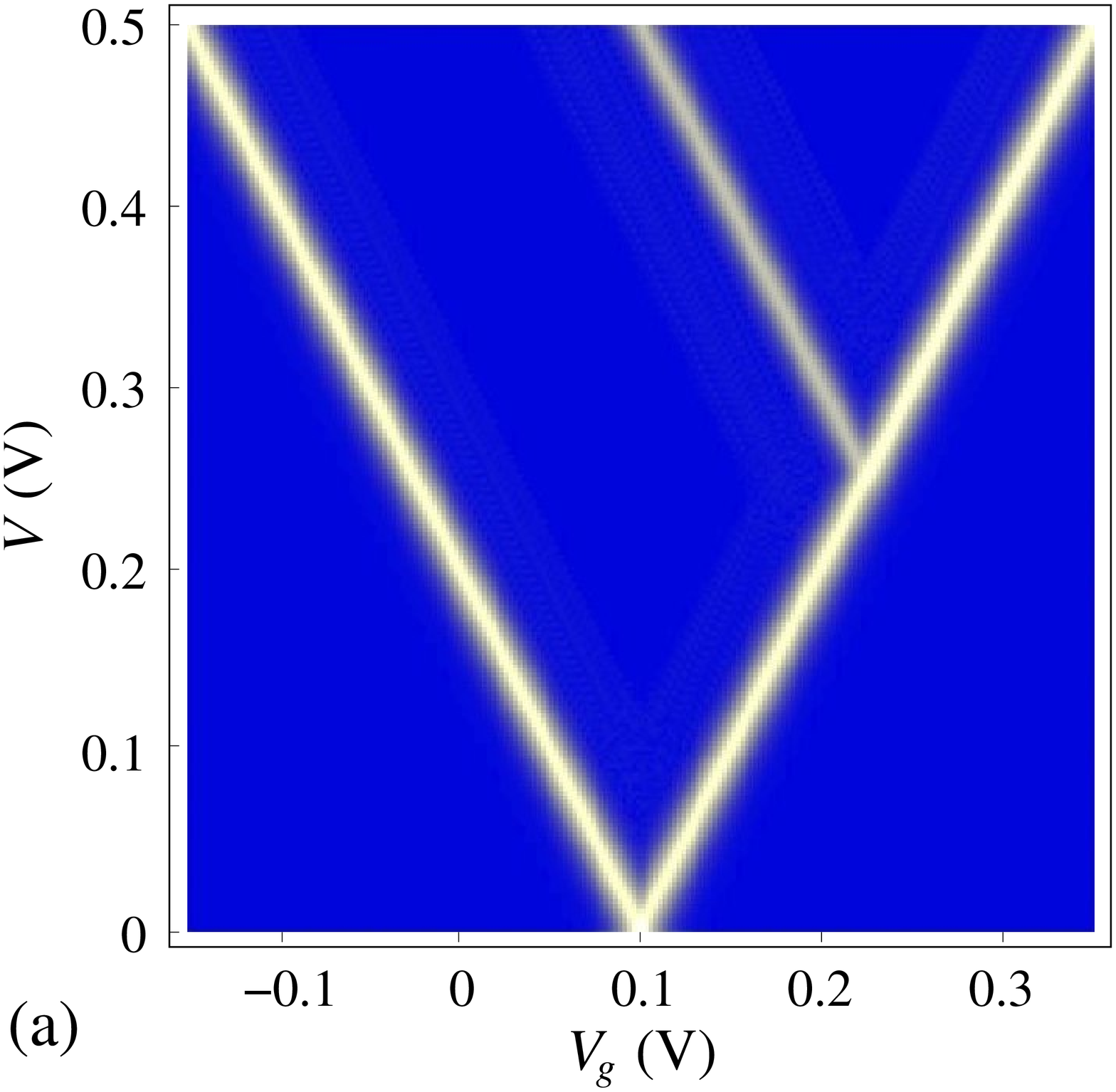}
\includegraphics[width=1.60in]{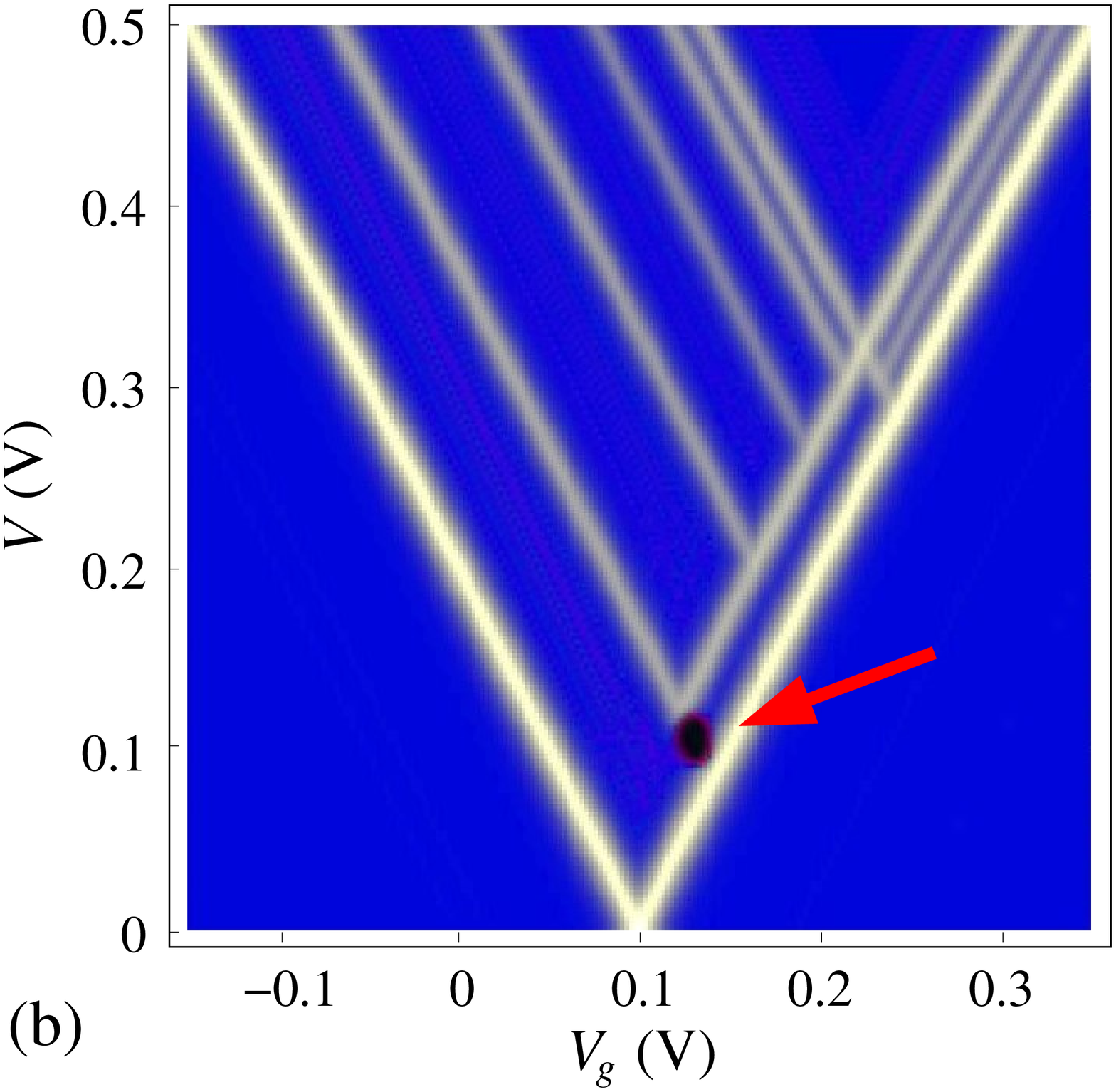}}
\vspace{1ex}
\centerline{\includegraphics[width=2.80in,clip]{fig1c.eps}}
\caption{(Color online) $dI/dV$ as a function of gate voltage $V_g$
and bias voltage $V$. Yellow (bright) colors denote positive values of the
differential conductance, dark red (dark gray) colors denote negative values.
In (a) results for an isotropic molecule are
shown, the parameters are $T=0.002$, $S=2$, $\epsilon=0.2$, $U=1$, $J=0.1$,
$K_2=B=0$. Energies are here and in the following given in electron volts.
Coulomb blockade is found to the left and right of the V-shaped region,
whereas within this region a nonzero steady-state current is flowing. The
satelite line results from the exchange splitting of energy levels for
$n=1$. In (b) an anisotropy of $K_2=0.04$ has been assumed, leading to a
complex splitting of the $dI/dV$ peaks. We find one peak
with NDC (arrow).
(c) Occupation, steady-state current in units of
$e/\tau_0$, $dI/dV$ in units of $e/\tau_0$ per volt,
and probabilities of various states (dashed) for a cut through
the NDC region in (b) at the lower temperature $T=0.001$.
The arrow indicates the plateau with reduced current.
Inset: Molecular energy levels of states with $n=0$
(black bars) and $n=1$ (red/gray bars with circles)
and magnetic quantum numbers $m$.}
\label{fig.fine}
\end{figure}

\section{Results}

As noted above, the anisotropy partially lifts the
degeneracy of molecular levels. The resulting fine structure of the peaks in
$dI/dV$ is shown in Fig.~\ref{fig.fine}(a) for vanishing anisotropy and in
Fig.~\ref{fig.fine}(b) for $K_2=0.04$. In the latter case
one of the peaks exhibits NDC. To
elucidate this effect we plot in Fig.~\ref{fig.fine}(c) the current,
differential conductance, and relevant occupation probabilities at
constant gate voltage at a lower temperature.

At low bias voltage only the degenerate ground states
$|1,5/2\rangle$ and $|1,-5/2\rangle$
are occupied and no current is flowing. For increasing
bias first the transitions to $|0,\pm 2\rangle$ become possible, cf.\ the
energy-level scheme in the inset in Fig.~\ref{fig.fine}(c).
The current then increases to a
plateau. On this plateau the four states have equal probability and
Eq.~(\ref{C1.I5}) yields $|\langle I^{\mathrm{L}}\rangle| = \frac{e}{2}\,
R^{\mathrm{L}}_{|1,\pm5/2\rangle \to |0,\pm 2\rangle}$. Next, the
transitions from $|0,\pm 2\rangle$ to $|1,\pm 3/2\rangle^-$ become active
and the system reaches a new plateau, denoted by an arrow in
Fig.~\ref{fig.fine}(c), with
\be
|\langle I^{\mathrm{L}}\rangle| = \frac{e}{3}\, \big( R^{\mathrm{L}}_{|1,\pm
5/2\rangle\to |0,\pm 2\rangle} + R^{\mathrm{L}}_{|1,\pm 3/2\rangle^-\to
|0,\pm 2\rangle} \big) .
\ee
Since for these particular parameters
$R^{\mathrm{L}}_{|1,\pm 3/2\rangle^-\to
|0,\pm 2\rangle}/R^{\mathrm{L}}_{|1,\pm 5/2\rangle\to |0,\pm 2\rangle} < 1/2$
due to the coefficients in Eq.~(\ref{C2.mixst4}),
the current is \emph{smaller} than on the first plateau, leading to NDC in the
crossover region. The transitions between plateaus are
rounded due to the nonzero temperature. This mechanism is
different from the one for $\mathrm{Mn}_{12}$ complexes\cite{HGF05}---in
our case there is no blocking state in which the molecule becomes trapped
due to suppressed outgoing transition rates. On the contrary, the
probabilities $P^i$ are equal for all accessable states.

\begin{figure}[t]
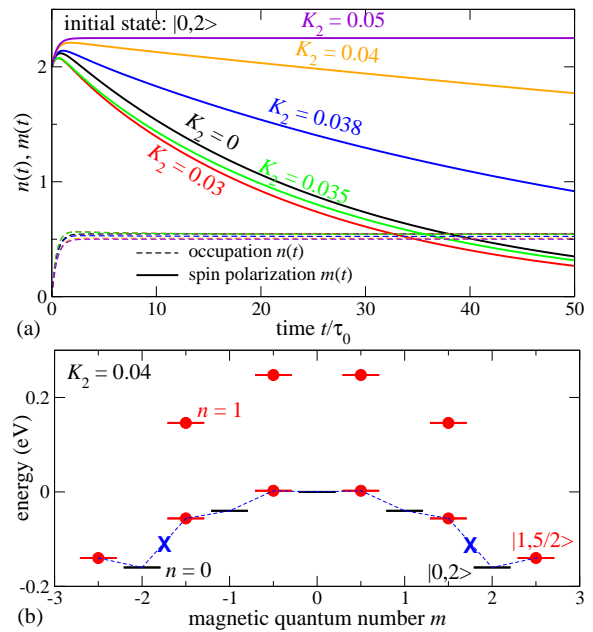

\centerline{\includegraphics[width=3.00in,clip]{fig2a.eps}}
\centerline{\includegraphics[width=3.00in,clip]{fig2b.eps}}
\caption{(Color online) (a) Occupation $n(t)$ and spin polarization
$m(t)$ for bias $V=0.2\,\mathrm{V}$, and $T=0.002$, $S=2$,
$\epsilon=0.12$, $U=1$, $J=0.1$, and $B=0$ for various
anisotropies $K_2$. In this and the following figures we absorp the gate
voltage into $\epsilon$.
The molecule is in state $|0,2\rangle$ at time $t=0$.
(b) Molecular energy levels for states with $n=0$ (black bars) and $n=1$
(red/gray bars with circles) for the parameters from (a) and $K_2=0.04$. Active
transitions are denoted by dashed lines. The crosses denote the
transitions that
next become thermally suppressed for larger $K_2$ or smaller bias $V$.}
\label{fig.relax}
\end{figure}

We now turn to the relaxation of the molecular spin. For a molecule prepared
in state $|0,2\rangle$ at time $t=0$, Fig.~\ref{fig.relax}(a) shows the
time dependence of occupation and spin polarization. The
occupation approaches the constant value $\langle n\rangle$ on the timescale
$\tau_0$, the typical time for a single tunneling event.

The spin polarization shows a quite different behavior with two distinct
timescales. Initially, $m(t)$ approaches a ``quasi-steady'' state on the
timescale $\tau_0$, which in this case has \emph{higher} polarization since
the state $|1,5/2\rangle$ has significant weight. Then, $m(t)$ decays to
zero more slowly. This decay is slow because the molecule must pass through
several intermediate states to reach a state with opposite spin
polarization, essentially performing a one-dimensional random walk, as
indicated in the level scheme, Fig.~\ref{fig.relax}(b). Figure
\ref{fig.relax}(a) shows that the spin relaxation initially becomes slightly
\emph{faster} for increasing $K_2$. This is due to the change of matrix elements
$C_{ij}^\sigma$ with $K_2$. For anisotropies
$K_2\gtrsim 0.04$ the decay becomes very slow since two of the transitions
needed to reverse the spin, denoted by the crosses in
Fig.~\ref{fig.relax}(b), become higher in energy than $eV/2$ and are thus
forbidden for $T\to 0$ and thermally activated for $T>0$.

\begin{figure}[t]
\centerline{\includegraphics[width=3.20in,clip]{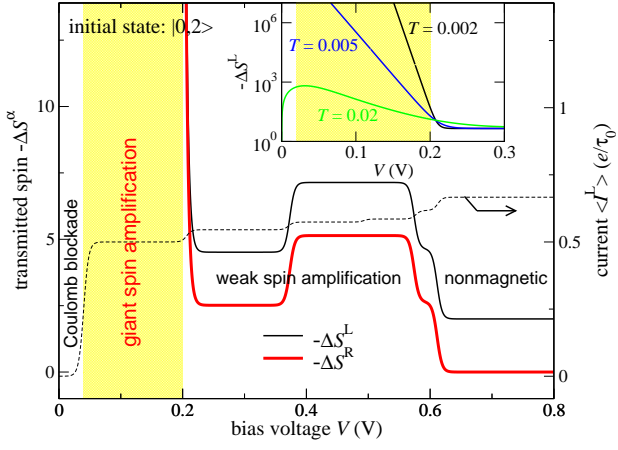}}
\caption{(Color online)
Total transmitted spin through the left and right leads for a
magnetic molecule prepared in the $n=0$ maximum spin state $|0,2\rangle$, as
functions of bias voltage. The parameters are as in
Fig.~\protect\ref{fig.relax} with $K_2=0.04$.
The steady-state current is also shown. The inset shows the spin transmitted
through the left lead on a logarithmic scale as a function of bias voltage
for three different temperatures.\protect\cite{footnote:smallbias}}
\label{fig.amplif}
\end{figure}

Importantly, in the regime of thermally activated (slow) spin relaxation a
sizeable steady-state current is flowing since transitions between the
states $|0,\pm 2\rangle$ and $|1,\pm 5/2\rangle$ are still possible, see
Fig.~\ref{fig.relax}(b). As noted above, this should lead to a nonzero total
transmitted spin, Eq.~(\ref{C1.DS2}). This is indeed found in
Fig.~\ref{fig.amplif}. Varying the bias voltage we observe four regimes:

(i) For small bias we are in the Coulomb blockade regime with very small
steady-state current. However, we find that the total transmitted spin is
exponentially \emph{large} in a characteristic energy barrier $\Delta E$ over
temperature. In our example $\Delta E$ is the difference between the energy
$\epsilon^-(1,3/2)-\epsilon(0,2)$ necessary for a \emph{spin-down} electron to
tunnel in and the available energy $eV/2$ of an incoming electron, which is
smaller. While all transitions
are thermally activated, the ones necessary to relax the spin have a much
higher energy than the transition between $|0,2\rangle$ and $|1,5/2\rangle$,
which dominates the current. However, the transmission takes exponentially
long since this transition is also thermally activated.

(ii) For larger bias we find the most interesting regime. The spin
relaxation rate is still small while the current is large. The bias is too
small to overcome the energy barrier between spin up and down, see
Fig.~\ref{fig.relax}(b). If the systems starts in state $|0,2\rangle$, the only
transition with large rate is to $|1,5/2\rangle$; the transitions to
$|1,3/2\rangle^\pm$ are thermally suppressed in this regime [note that they are
higher in energy in Fig.~\ref{fig.relax}(b)]. Thus a \emph{spin-up} electron has
to enter since the process increases the total spin from $2$ to $5/2$. This
electron can leave the molecule through the other lead, returning it to the
intial state. On the other hand, a second electron cannot enter the molecule
since this would cost a high Coulomb energy $U$. Therefore, the current if fully
spin polarized until a transition to $|1,3/2\rangle^-$ occurs. This thermally
suppressed transition happens with a small rate proportional to $\exp(-\Delta
E/T)$, where $\Delta E=\epsilon^-(1,3/2)-\epsilon(0,2)-eV/2$, as above.
Consequently,
the current is spin-polarized for an
exponentially long (in $\Delta E/T$)
time leading to an exponentially large transmitted spin
$|\Delta S^\alpha|$ [the negative sign in Fig.~\ref{fig.amplif} can be
understood by considering the transition rates, Eq.~(\ref{C1.R4}), in
detail]. The exponential dependence of the transmitted spin on $\Delta E/T$ is
clearly seen in the inset in Fig.~\ref{fig.amplif}.
The average time $T_s^\alpha \equiv {1}/{|I_s^\alpha(t=0)|}$ for
one unit of spin to be transmitted is of the order of
$\tau_0$.\cite{footnote:instant} If
$T_s^{\mathrm{R}}$ is short compared to the spin relaxation time in the
leads large opposite magnetizations will be accumulated in the leads,
similarly to the effect of spin injection studied intensively in recent
years.\cite{Sch05} This
effect is induced by the breaking of spin symmetry at $t=0$ only through the
polarization of a single quantum spin, and can thus be described as
\emph{giant spin amplification}. It is a promising method to \emph{read out}
the spin information. The magnetization in the leads could be
detected with a pickup coil or by the
magneto-optical Kerr effect.\cite{BlH94,KMG04}
A strong amplification mechanism could also facilitate the reliable
transfer of spin between individual molecules in a device.

(iii) Further increase of the bias leads to a regime where the transmitted
spin is nonzero but not exponentially enhanced. Here, both spin relaxation
rate and current are large. The bias is large enough to overcome the energy
barrier. Figure \ref{fig.amplif} shows that the spin transmitted through the
left lead is (in absolute value) larger by two than through the right lead.
The reason is that the electrons flow from right to left and that the spin
of the initial state has to leave the molecule for $t\to\infty$.

(iv) At large bias \emph{all} transitions between states with $n=0, 1$ are
possible. The spin transmitted through the right, incoming lead, $\Delta
S^{\mathrm{R}}$, essentially vanishes. The initial spin leaves the molecule
through the left lead, leading to $\Delta S^{\mathrm{L}}\cong -2$. This is
the nonmagnetic regime.

The pattern seen in Fig.~\ref{fig.amplif} is robust under change of anisotropy
$K_2$. For $K_2=0$ the regime of giant spin amplification is absent but for any
$K_2>0$ it exists at sufficiently low temperatures since the only requirement
is that the energy of state $|1,5/2\rangle$ is larger than the one of
$|1,3/2\rangle^-$.

Finally, for molecular memory application one also needs to \emph{write} the
information, i.e., to switch the molecule to a predetermined state. An
obvious idea is to apply a magnetic field to a molecule attached to the
nonmagnetic leads considered so far. However, applying a field at zero bias
does not work since all transitions remain thermally suppressed so that
relaxation to the spin-polarized steady state is exponentially slow.

\begin{figure}[t]
\centerline{\includegraphics[width=3.00in,clip]{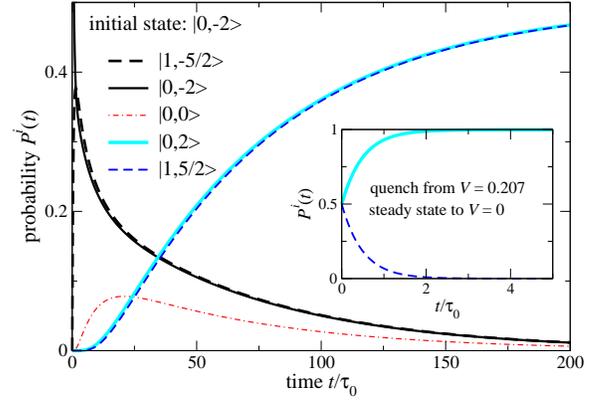}}
\caption{(Color online) Probabilities of molecular states after initial
preparation in state $|0,-2\rangle$ as functions of time at the bias voltage
$V=0.207\,\mathrm{V}$.  The probabilities of states with extremal spin
polarization and of the representative intermediate state $|0,0\rangle$ are
shown. A magnetic field ($B=0.002$) leads to spin relaxation towards larger
spin values. The parameters are $T=0.0002$ (lower than before), $K_2=0.04$
and otherwise as in Fig.~\protect\ref{fig.relax}(a). Inset: Probabilities
after the bias voltage is suddenly switched off in the steady state reached
for $V=0.207\,\mathrm{V}$.}
\label{fig.write}
\end{figure}

We find that reliable switching requires a \emph{two-step scheme}: First one
applies a magnetic field, which \emph{tilts} the energy levels in
Fig.~\ref{fig.relax}(b), and a bias voltage that is just large enough to
allow transitions in the desired direction but not in the opposite one at
low temperatures. Since the Zeeman energy $B$ is small this requires fine
tuning of $V$ on the scale $B/e$ and cooling to $T\ll B$. Figure
\ref{fig.write} shows the change of probabilities for all states with $n=0,
1$ if one starts with $|0,-2\rangle$ and applies a positive $B$ field to
switch the molecule to spin up. The Zeeman energy is chosen as
$2\,\mathrm{meV}$. The molecule crosses over from $|0,-2\rangle$ to a steady
state essentially consisting of $|0,2\rangle$ and $|1,5/2\rangle$ on a
timescale of the same order and of the same origin as the spin
relaxation times, see Fig.~\ref{fig.relax}. Note that in a molecular circuit
one could apply a magnetic field to many molecules and address a specific
one with the bias voltage. In the second step the bias is switched off. The
inset of Fig.~\ref{fig.write} shows that the system then relaxes towards the
target state $|0,2\rangle$ on the timescale $\tau_0$.

\begin{figure}[t]
\centerline{\includegraphics[width=3.20in,clip]{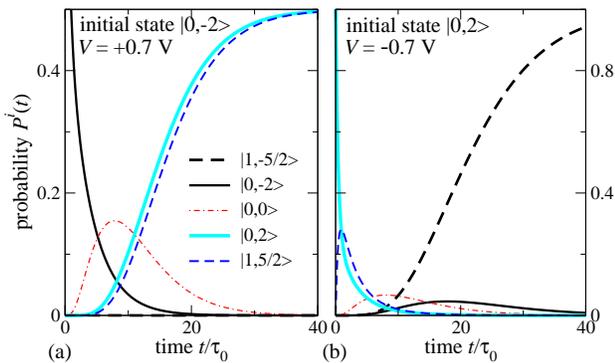}}
\caption{(Color online) Probabilities of molecular states after initial
preparation in state (a) $|0,-2\rangle$ and (b) $|0,2\rangle$ as functions
of time at the bias voltage (a) $V=0.7\,\mathrm{V}$ and (b)
$-0.7\,\mathrm{V}$.
The right lead is ferromagnetic with a ratio of densities of states
of $D^{\mathrm{R}}_\downarrow/D^{\mathrm{R}}_\uparrow = 0.001$. There is no
applied magnetic field. The other parameters are as in
Fig.~\protect\ref{fig.write} except for the \emph{much higher} temperature
$T=0.02$.}
\label{fig.fmwrite}
\end{figure}

This writing scheme is problematic because it requires to reverse a large
magnetic field between switching events, which is a very slow process. It also
requires very low temperatures. We can overcome these problems by using one
ferromagnetic and one nonmagnetic lead. The ferromagnetic lead R is modelled by
different densities of states $D^{\mathrm{R}}_\sigma$ for
$\sigma=\uparrow,\downarrow$ electrons. We set
$D^{\mathrm{R}}_\downarrow/D^{\mathrm{R}}_\uparrow = 0.001$ so that the lead is
essentially a half-metallic ferromagnet such as $\mathrm{NiMnSb}$ and
$\mathrm{CrO}_2$. For less complete spin polarization in the lead the
reliability of the switching degrades. We note that spin injection from a
ferromagnet is not trivial. However, in recent years significant progress has
been made.\cite{Sch05} Nevertheless, the most difficult task in
implementing this writing scheme probably is to achieve a high degree of spin
polarization of tunneling electrons.

Figure \ref{fig.fmwrite} shows
that one can switch the spin polarization in \emph{both} directions on the
timescale of typical spin relaxation times by applying a bias voltage in
\emph{vanishing} magnetic field.
For $V>0$, Fig.~\ref{fig.fmwrite}(a), the electrons flow from right to left.
Since $D^{\mathrm{R}}_\downarrow$ is small, nearly all electrons have spin
up. Due to exchange scattering between electron spin and local spin,
the latter is \emph{increased}. Figure \ref{fig.fmwrite}(a) shows that all
states except $|0,2\rangle$ and $|1,5/2\rangle$, which have the largest positive
magnetization, die out. By switching the bias voltage to $V=0$ as in
Fig.~\ref{fig.write} we can then make the molecule relax rapidly towards the
unique state $|0,2\rangle$ (not shown). For $V<0$,
Fig.~\ref{fig.fmwrite}(b), the electrons flow from left to right. Spin-down
electrons are essentially trapped on the molecule until they perform a spin
exchange with the local spin, which \emph{decreases} the local spin.
As the result, the state $|1,-5/2\rangle$ becomes populated at the expense of
all other states.
Importantly, no fine tuning of the bias voltage is required. Furthermore,
the temperature need not be small---Fig.~\ref{fig.fmwrite} is calculated
with $T$ close to room temperature. Both properties are desirable for
molecular-electronics applications.

If one lead is ferromagnetic, the current is generally spin-polarized even
in the steady state. Then the total transmitted spin,
Eq.~(\ref{C1.DS2}), diverges. However, we can define the \emph{excess}
transmitted spin in a certain state relative to the steady state,
$\Delta S^\alpha[\mathbf{P}] = \int_0^\infty dt\, [I_s^\alpha(t,\mathbf{P})
- \langle I_s^\alpha\rangle]$. This quantity exhibits spin
amplification as in the case of nonmagnetic leads.

\section{Conclusions}

We have studied the inelastic charge and spin
transport through a magnetically anisotropic molecule weakly coupled to
metallic leads. The three processes crucial for molecular memory
applications---writing, storing, and reading information---can be
implemented in such a device. The information storage is affected by spin
relaxation, which can be very slow for large easy-axis anisotropy. Also due
to the anisotropy, application of a bias voltage to a molecule in a
spin-polarized state can lead to the transfer of a large amount of spin or
magnetic moment from one lead to the other. This transmitted spin increases
exponentially for low temperatures.
We propose that this giant spin amplification
could be used to read out the spin information. Finally, we also propose a
scheme to write the information, which does not require a
magnetic field but uses one ferromagnetic lead.

\acknowledgments

We would like to thank J. Koch and F. von Oppen for helpful discussions.
Support by the Deutsche For\-schungs\-ge\-mein\-schaft through
Sfb 658 is gratefully acknowledged.


\begin{thebibliography}{50}
\bibitem{footnote:smallbias}For small bias, $V\to 0$, the transmitted spin for
the two lower temperatures should also decrease again, like it does for
$T=0.02$. No results are plotted in this regime due to round-off errors in the
calculation of the steady-state probabilities.
\bibitem{footnote:instant}More precisely, the total transmitted spin for a long
time interval $\Delta t\gg\tau_0$ is of the order of $\Delta t/\tau_0$. The
tunneling of a single electron is instantaneous in our model.
\end{thebibliography}
\end{document}